\documentclass[onecollarge,natbib]{svjour2}
\bibpunct{[}{]}{;}{n}{}{,} 
\smartqed  
\usepackage{graphicx}
\journalname{Few-Body Systems}
\begin{document}

\title{Jefferson Lab Science
}
\subtitle{Present and Future}


\author{R. ~D.~McKeown         
}


\institute{R.~D.~McKeown \at
              Jefferson Lab, Newport News, VA, USA \\
              Physics Department, College of William and Mary, Williamsburg, VA, USA \\
              Tel.: 757-269-6481\\
              \email{bmck@jlab.org}           
}

\date{Received: date / Accepted: date}

\maketitle

\begin{abstract}
The Continuous
Electron Beam Accelerator Facility (CEBAF) and associated experimental equipment at Jefferson Lab comprise a unique facility for experimental nuclear physics. This facility is presently being upgraded, which will enable a new experimental program with substantial discovery potential to address important topics in nuclear, hadronic, and electroweak physics. Further in the future, it is envisioned that the Laboratory will evolve into an electron-ion colliding beam facility.
\end{abstract}

\section{Introduction}
\label{intro}
Since 1995, the CEBAF facility at Jefferson Laboratory has operated
high-duty factor (continuous) beams of up to 6~GeV electrons incident on three
experimental halls (denoted A, B, and C), each with a unique set of experimental
equipment. The development of advanced GaAs photoemission sources has enabled high
quality polarized beam with polarizations up to 85\% at beam currents up to 180~$\mu$A. The facility
serves an international scientific user community of over 1300
scientists, and to date 178 experiments have been completed.
The first decade of scientific results from Jefferson Lab has recently been summarized in an extensive review \cite{decade}. This remarkable record of scientific productivity, and the prospects for further scientific advances in this field,  have motivated a major upgrade of CEBAF to 12 GeV electron beam energy along with substantial new experimental equipment.

\section{The 12 GeV Upgrade Project}
\label{sec:1}
\begin{figure}
\begin{center}
\includegraphics[width=6in]{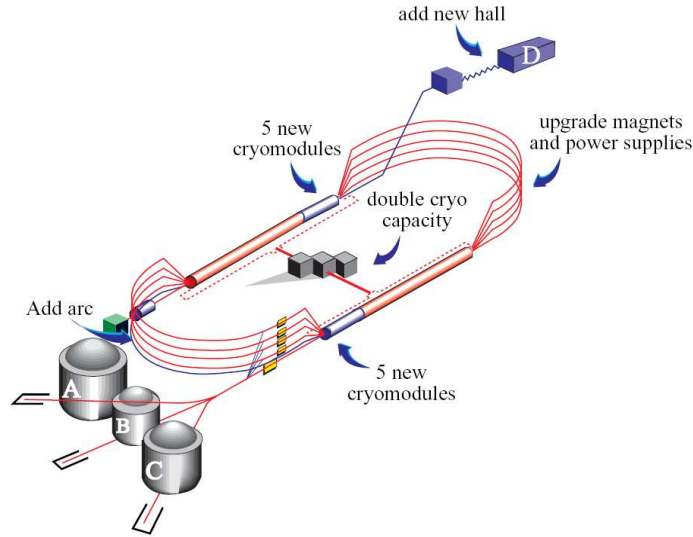}
\end{center}
\caption{\label{Fig:Upgrade}Jefferson Lab 12 GeV upgrade concept.}
\end{figure}

The upgrade of CEBAF and associated experimental equipment at Jefferson Lab is presently underway, with completion expected in 2017.  The upgraded facility, shown in Fig.~\ref{Fig:Upgrade} will accelerate electron beams to 11 GeV for experiments in the existing Halls A, B and C.  In addition, a 12~GeV beam can be provided to a new experimental Hall D to generate a 9 GeV tagged photon beam, enabling a powerful program of meson spectroscopy.  The facility will capable of delivering beam to any 3 of the 4 halls simultaneously.

The accelerator portion of the upgrade project is essentially complete, and commissioning is underway. In early 2014, accelerator operators sent electrons around the CEBAF accelerator and achieved full upgrade-energy acceleration of 2.2 GeV in one pass.  Subsequently, the CEBAF accelerator delivered beam into a target in Hall A, recording the first data of the 12 GeV era.  The operations staff then tuned up a 3-pass beam, resulting in 6.11 GeV electrons at 2 nanoAmps average current for more than an hour. On May 3, 2014 the first 5.5 pass beam, with energy of 6.18 GeV, was delivered to the front section of the beamline to Hall D, thus demonstrating that all 5.5 passes of the accelerator were functional. Shortly thereafter, the machine delivered its highest-energy beams ever, 10.5 GeV, through the entire accelerator and into the Hall D Tagger Facility, which converts CEBAF's electron beam into photons that will be used for experiments in Hall D. Beam commissioning will continue in the Fall of 2014 with the aim of establishing 12 GeV beam energy and delivering simultaneous beam to Hall A and D.

The GlueX experimental apparatus in Hall D is essentially complete and ready for commissioning this fall. CLAS12 in Hall B and Super High Momentum Spectrometer (SHMS)in Hall C are still in the construction phase. While detector construction is well advanced, the superconducting magnets for these projects are still under construction at the vendors.
Current projections indicate that  CLAS12 and the SHMS spectrometer will be ready for beam commissioning in summer 2016.  

\section{The 12 GeV Science Program}

The physics program to be addressed with the Jefferson Lab 12~GeV upgrade has been developed in collaboration with the user
community and with the guidance of the Program Advisory Committee. There are presently 70 approved experiments, which will likely require at least a decade to execute.

The major science questions to be addressed with the upgraded facility include:
\begin{itemize}
  \item What is the role of gluonic excitations in the spectroscopy of light mesons? Can these excitations elucidate the origin of quark confinement?
  \item Where is the missing spin in the nucleon? Is there a significant contribution from valence quark orbital angular momentum?
  \item Can we reveal a novel landscape of nucleon substructure through measurements of new multidimensional distribution functions?
  \item What is the relation between short-range N-N correlations and the partonic structure of nuclei?
  \item Can we discover evidence for physics beyond the standard model of particle physics?
\end{itemize}
A detailed exposition of the science opportunities at the upgraded facility has recently been published \cite{12WP}. Here we briefly present some highlights.

\subsection{Meson Spectroscopy}
 
 Excitation of the gluonic field in a quark-antiquark system can lead to mesonic states with exotic quantum numbers
 ($J^{PC}= 0^{+-}, 1^{-+}, 2^{+-}$) that cannot be
described by states with only quark-antiquark degrees of freedom. These states and their
properties have recently been studied in detail using lattice QCD methods
\cite{LQCD, LQCD2}. Chiral extrapolation of these lattice calculations to the physical pion mass indicate that the exotic mesons will indeed be present in the mass range 2-2.5~GeV.

A major new experiment, known as GlueX, has been constructed and sited in Hall D. The main goal of the
GlueX experiment is to search for exotic mesons produced via
photoproduction on the nucleon.  Linearly polarized photons will be produced upstream of Hall D by a
coherent bremmstrahlung process using a thin ($\sim 20$ micron) diamond wafer.
The scattered electrons from 8.5-9~GeV bremmstrahlung photons will be tagged with
scintillator detectors following a bending magnet, yielding a tagged
photon resolution of 0.2\% with fluxes expected to reach $10^8$/s.

\subsection{Nucleon Structure}
\label{sec:3.2}

Since the initial discovery at SLAC in the 1960's, the partonic structure of the nucleon has been described by the one-dimensional parton distribution functions (PDF). The four parity conserving structure functions of the variable $x$ ($f_1$, $f_2$, $g_1$, and $g_2$), have been studied in great detail in many experiments over the last four decades. However, over the last 15 years it has been realized that these one dimensional distributions are not adequate to describe the nucleon and do not contain some essential physics needed to provide a complete picture. The orbital angular momentum of partons is a good example of a degree of freedom that is not exhibited in the standard 1-d PDFs. Recent experimental and theoretical studies indicate that a more complete description of the partonic structure of the nucleon is realizable through new multi-dimensional distributions: Generalized Parton Distributions (GPD) and Transverse Momentum Dependent (TMD) distributions.

GPDs contain information on the correlation between the quark/gluon transverse position in the nucleon and its longitudinal momentum They can be accessed in exclusive scattering processes at large $Q^2$: deeply virtual Compton scattering (DVCS) and deep virtual meson production (DVMP).  GPDs offer a path to a full 3-dimensional exploration of nucleon structure, in transverse position and longitudinal momentum space, enabling spatial tomography of the nucleon. The new CLAS12 apparatus being constructed for Hall B as part of the 12 GeV upgrade is specifically designed to study these processes.

Transverse momentum dependent distributions (TMDs) contain information on the quark/gluon intrinsic motion in a nucleon, and on the correlations between the transverse momentum of the quark and the quark/nucleon spins.  TMDs offer a unique opportunity for a momentum tomography of the nucleon, and can be measured in Semi-Inclusive Deep Inelastic Scattering (SIDIS), in which the nucleon is no longer intact and one of the outgoing hadrons is detected. SIDIS will be studied in Hall C with high resolution spectrometers, in Hall B with CLAS12, and in Hall A with the new Super Bigbite Spectrometer (SBS). A major new capability for SIDIS is proposed in the Solenoidal Large Intensity Device (SoLID) to be sited in Hall A. This project is still under development, but would offer exceptional capability for mapping out the TMDs with the 12 GeV beam at JLab.

\section{Quarks in Nuclei}

The Jefferson Lab 12 GeV Upgrade will both study the QCD structure of nuclei and use the nucleus as a laboratory to study QCD.  The new facility will enable investigations of a number of the most fundamental questions in modern nuclear physics.

The nature of the nucleon-nucleon (NN) relative wave function at short distances is fundamental to the origin of the nuclear force and to the properties of nuclei. It is not known if this system can be described only in terms of nucleons and mesons, or whether quarks and gluons are necessary for its description. Recent studies indicate that modification of the nuclear parton distributions, or the "EMC effect", is related to short-range NN correlations in nuclei \cite{EMC_SRC}. Further experimental studies to explore these important issues will be possible with the 12 GeV CEBAF and new experimental equipment.

QCD also suggests the existence of novel phenomena in nuclear physics.  The nuclear medium provides mechanisms for filtering quantum states and studying their spacetime evolution. Studying the hadronization of a struck quark in different nuclei affords a unique method for elucidating this process. The formation of small color singlet configurations leads to the novel process known as color transparency. The increased kinematic range of the 12 GeV CEBAF will offer new opportunities to study these and other related topics.

A recent Jefferson Lab experiment, PREX, has demonstrated a new method to determine the neutron radius of a heavy nucleus like $^{208}$Pb \cite{PREX}. The precise measurements of the charge distribution of nuclei in elastic electron scattering provide stringent constraints on the distribution of protons in nuclei. However, the distribution of neutrons is more difficult to study and is also quite important for predicting the properties of neutron stars. The weak charge of the neutron is -1, whereas the weak charge of the proton is $1-4\sin^2 \theta_W \ll 1$. Thus, the measurement of parity violating asymmetries in elastic electron scattering from nuclei is sensitive to the neutron distribution and can be used to constrain the neutron radius. Future studies with higher precision are planned for the Jefferson Lab 12 GeV program.

\section{Beyond the Standard Model}

Exciting new opportunities to search for new physics beyond the Standard Model will become possible at Jefferson Lab in the 12 GeV era. The very precise measurements of parity violating asymmetries to study the strange form factors in elastic electron-proton scattering have demonstrated that this technique has substantial potential for precision tests of the Standard Model. The strength of the neutral weak interaction is parameterized in the standard model by the weak mixing angle $\theta_W$. This parameter is very precisely determined at the $Z$ boson mass by $e^+$-$e^-$ collider experiments. The two best measurements (which differ by more than $2\sigma$) have uncertainties of 0.00029 and 0.00026, and can be combined to yield the average value $\sin^2 \theta_W =  0.23116  \pm 0.00013 $ \cite{pdg}. Radiative corrections associated with standard model physics predicts a ``running'' of this coupling to $\sin^2 \theta_W =  0.2388$ at $Q^2 =0$. Additional particles at high mass (larger than $M_Z$) would generally modify these radiative corrections, leading to a different value of $\sin^2 \theta_W $ at $Q^2=0$. Thus precise measurements of the neutral weak interaction at low $Q^2 \ll M_Z^2$ can reveal the presence of particles and forces not present in the standard model.

A major new experiment to study parity-violation in elastic electron-proton scattering at low $Q^2$ was recently completed in Hall C at Jefferson Lab \cite{Qweak}.
The data from this experiment, known as $Q_{weak}$, are under analysis. In addition,
there are presently 2 new proposals to perform parity violation measurements at the
upgraded CEBAF. One would use the proposed solenoidal magnetic
spectrometer system (SOLID) to study parity-violating deep inelastic
scattering \cite{SOLID}. The other proposal involves the
construction of a novel dedicated toroidal spectrometer to study
parity-violating M{\o}ller scattering \cite{Moller}. Both
experiments will require construction of substantial new
experimental equipment (beyond the scope of the present upgrade
project) and are proposed to be sited in experimental Hall A.

Heavy photons, called A's, are new hypothesized massive vector bosons that have a small coupling to electrically charged matter, including electrons.  The existence of an A' is theoretically natural and could explain the discrepancy between the measured and observed anomalous magnetic moment of the muon \cite{g-2} and several intriguing dark matter-related anomalies.  New electron fixed-target experiments proposed at Jefferson Lab, with its high-quality and high-luminosity electron beams, present a unique and powerful probe for A's.  These experiments include the A' Experiment (APEX)\cite{APEX}, the Heavy Photon Search (HPS)\cite{HPS}, and Detecting A Resonance Kinematically with Electrons Incident on a Gaseous Hydrogen Target (Dark Light)\cite{DARKLIGHT}.

\section{Electron Ion Collider (EIC)}

As discussed in \ref{sec:3.2} the upgraded CEBAF at Jefferson Lab will provide unprecedented capability to study nucleon structure in the valence region where Bjorken $x>0.1$. There remains a crucial need to study the nucleon at lower $x<0.1$. From measurements in the 1990's at the HERA facility, it is known that at low $x$ the parton dynamics is dominated by a large rise in the density of gluons. While HERA collided electrons and protons are large center of mass energies, it did not have polarized nucleon beams or nuclear beams, and the luminosity was constrained to be rather low ($\sim 10^{31}$~cm$^{-2}$~s$^{-1}$. This limited luminosity is not sufficient to study nucleon tomography with GPD's and TMD's as will be possible at larger $x$ with the 12 GeV CEBAF. Recently, a white paper that addresses the scientific potential of the envisioned EIC facility has been produced \cite{EICWP}. The community reached a consensus on the basic scientific requirements for such a facility:
\begin{itemize}
  \item Highly polarized ($\sim 70$\%) electron and nucleon beams
  \item Ion beams from deuteron to the heaviest nuclei (Uranium or Lead)
  \item Variable center of mass energies from $\sim 20 -  \sim 100$~ GeV, upgradable to $\sim 150$~ GeV
  \item High collision luminosity $\sim 10^{33-34} {\rm cm}^{-2} {\rm s}^{-1}$.
\end{itemize}
Such a facility would capitalize on the powerful new experimental techniques for exploring nucleon structure that are being developed for the 12 GeV JLab program, and apply them to the low $x$ region where the dynamics is dominated by the gluons. It is widely perceived that addressing this kinematic regime with high luminosity and fully polarized beams is necessary to complete our understanding of the basic partonic structure of the nucleon. An EIC is viewed as a natural extension of the capabilities of the Jefferson Lab 12 GeV upgrade, the RHIC spin program, and the COMPASS experiment at CERN.

\begin{figure}
\begin{center}
\includegraphics[width=3.5 in]{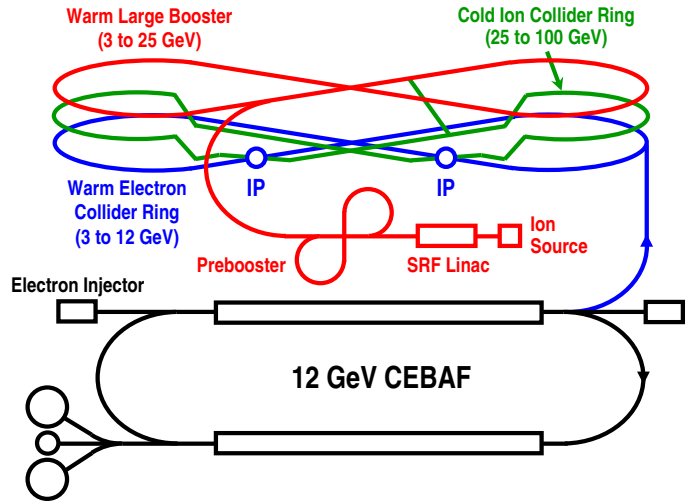}
\end{center}
\caption{\label{fig:MEIC} A schematic drawing of MEIC.}
\end{figure}

The Accelerator and Physics Divisions at Jefferson Lab have been working for several years on a novel design for an EIC that would utilize the 12 GeV CEBAF as an injector to a collider facility \cite{MEIC12}. As shown in Figure~\ref{fig:MEIC}, the storage rings would be in a "figure 8" layout to mitigate the effects of depolarizing resonances and facilitate high beam polarization. A medium energy version, MEIC, is envisioned that would collide 12 GeV electrons with 100 GeV protons. This could be upgraded to the full EIC facility with 12 GeV electrons colliding with 250 GeV protons.




\end{document}